\input phyzzx.tex
\tolerance=1000
\voffset=-0.0cm
\hoffset=0.7cm
\sequentialequations
\def\rl{\rightline}

\def\t1{{\tilde 1}}

\def\SUSY{supersymmetry }

\def\t{\theta}

\REF{\OGM}{M. Dine, W. Fischler and M. Srednicki, Nucl. Phys. {\bf B189} (1981) 575; S. Dimopoulos and S. Raby, Nucl. Phys. {\bf B192} (1982) 353; M. Dine and W. Fischler, Phys. Lett. {\bf B110} (1982) 227; Nucl. Phys. {\bf B204} (1982) 346; C. Nappi and B. Ovrut, Phys. Lett. {\bf B113} (1982) 175; L. Alvarez-Gaume,
M. Claudson and M. Wise, Nucl. Phys. {\bf B207} (1982) 96; S. Dimopoulos and S. Raby, Nucl. Phys. {\bf B219} (1983) 479.}
\REF{\GM}{M. Dine and A. E. Nelson, Phys. Rev. {\bf D48} (1993) 1277, [arXiv:hep-ph/9303230]; M. Dine, A. E. Nelson and Y. Shirman, Phys. Rev. {\bf D51} (1995) 1362, 
[arXiv:hep-ph/9408384]; M. Dine, A. E. Nelson, Y. Nir and Y. Shirman, Phys. Rev. {\bf D53} (1996) 2658, [arXiv:hep-ph/9507378].} 
\REF{\GUI}{G. F. Guidice and R. Rattazzi, Phys. Rept. {\bf 322} (1999) 419, [arXiv:hep-ph/9801271].}
\REF{\SUSY}{Y. Shadmi and Y. Shirman, Rev. Mod. Phys. {\bf 72} (2000) 25, [arXiv:hep-th/9907225]; J. Terning, [arXiv:hep-th/0306119]; K. Intriligator and N. Seiberg, 
Class. Quant. Grav. {\bf 24} (2007) S741, [arXiv:hep-th/0702069].}
\REF{\ISS}{K. Intriligator, N. Seiberg and D. Shih, JHEP {\bf 0604} (2006) 021, [arXiv:hep-th/0602239].}
\REF{\DIN}{M. Dine, J. L. Feng and E. Silverstein, Phys. Rev. {\bf D74} (2006) 095012, [arXiv:hep-th/0608159].}
\REF{\RAY}{S. Ray Phys. Lett {\bf B642} (2006) 13, [arXiv:hep-th/0607172].}
\REF{\KIT}{R. Kitano, hep-ph/0606129; Phys. Lett. {\bf B641} (2006) 203, [arXiv:hep-ph/0607090].}
\REF{\DM}{M. Dine and J. Mason, Phys. Rev. {\bf D77} (2008) 016005, [arXiv:hep-ph/0608063].}
\REF{\KOO}{R. Kitano, H. Ooguri and Y. Ookouchi, Phys. Rev. {\bf D75} (2007) 045022, [arXiv:hep-ph/0612139].}
\REF{\MUR}{H. Murayama and Y. Nomura, Phys. Rev. Lett. {\bf 98} (2007) 151803, [arXiv:hep-ph/0612186].}
\REF{\OS}{O. Aharony and N. Seiberg, JHEP {\bf 0702} (2007) 054, [arXiv:hep-ph/0612308].}
\REF{\CST}{C. Csaki, Y. Shirman and J. Terning, JHEP {\bf 0702} (2007) 099, [arXiv:hep-ph/0612241].}
\REF{\SHI}{K. Intriligator, N. Seiberg and D. Shih, JHEP {\bf 0707} (200) 017, [arXiv:hep-th/0703281].}
\REF{\RSY}{D. Shih, JHEP {\bf 0802} (2008) 091, [arXiv:hep-ph/0703196].}
\REF{\GIV}{A. Giveon and D. Kutasov, Nucl. Phys. {\bf B796} (2008) 25, arXiv:0710.0894[hep-th].}
\REF{\ZKKS}{A. Giveon, A. Katz, Z. Komargodski and D. Shih, JHEP {\bf 0810} (2008) arXiv:0808.2901[hep-th].}
\REF{\KOM}{Z. Komargodski and D. Shih, JHEP {\bf 0904} (2009) 093, arXiv:0902.0030[hep-th].}
\REF{\ESUS}{E. Halyo, arXiv:0906.2127[hep-ph].}
\REF{\BGH}{I. Bena, E. Gorbatov, S. Hellerman, N. Seiberg and D. Shih, JHEP{\bf 0611} (2006) 088, [arXiv:hep-th/0608157].}
\REF{\FRA}{S. Franco, I. Garcia-Etxebarria and A.M. Uranga, JHEP {\bf 0701} (2007) 085, [arXiv:hep-th/0607218].}
\REF{\AMI}{A. Giveon and D. Kutasov, Nucl. Phys. {\bf B778} (2007) 129, [arXiv:hep-th/0703135]; JHEP {\bf 0802} (2008) 038, arXiv:0710.1833[hep-th].}
\REF{\ARG}{R. Argurio, M. Bertolini, S. Franco and S. Kachru, JHEP {\bf 0706} (2007) 017, [arXiv:hep-th/0703236].}
\REF{\BER}{M. Bertolini, S. Franco and S. Kachru, JHEP {\bf 0701} (2007) 017, [arXiv:hep-th/0610212].}
\REF{\OOG}{H. Ooguri and Y. Ookouchi, Phys. Lett. {\bf B641} (2006) 323, [arXiv:hep-th/0607183].}
\REF{\AGA}{M. Aganagic, C. Beem, J. Seo and C. Vafa, Nucl. Phys. {\bf B789} (2008) 382, [arXiv:hep-th/0610249].}
\REF{\ABK}{M. Aganagic, C. Beem ad S. Kachru, Nucl. Phys. {\bf B796} (2008) 1, arXiv:0709.4277[hep-th].}
\REF{\OKS}{O. Aharony, S. Kachru and E. Silverstein, Phys. Rev. {\bf D76} (2007) 126009, arXiv:0708.0493[hep-th].}
\REF{\FU}{S. Franco and A. M. Uranga, JHEP {\bf 0606} (2006) 031, [arXiv:hep-th/0604136].}
\REF{\AHN}{C. Ahn, Class. Quant. Grav. {\bf 24} (2007) 1359, [arXiv:hep-th/0608160]; Class. Quant. Grav. {\bf 24} (2007) 3603, [arXiv:hep-th/0702038].}
\REF{\KAW}{T. Kawano, H. Ooguri and Y. Ookouchi, Phys. Lett. {\bf B652} (2007) 40, arXiv:0704.1085[hep-th].}
\REF{\TAT}{R. Tatar and B. Wettenhall, Phys. Rev. {\bf D76} (2007) 126011, arXiv:0707.2712[hep-th].}
\REF{\VER}{M. Buican, D. Malyshev and H. Verlinde, JHEP {\bf 0806}(2008) 108, arXiv:0710.5519[hep-th].}
\REF{\ARG}{R. Argurio, M. Bertolini, G. Ferretti and A. Mariotti, arXiv:0906.0727[hep-th].}
\REF{\ORI}{A. Giveon, D. Kutasov, J. McOrist and A. B. Royston, Nucl. Phys. {\bf B822} (2009) 106, arXiv:0904.0459[hep-th].}
\REF{\LAST}{E. Halyo, arXiv:0906.2377[hep-th].}
\REF{\QUI}{F. Cachazo, S. Katz and C. Vafa, [arXiv:hep-th/0108120].}
\REF{\SUP}{E. Witten, Nucl. Phys. {\bf B507} (1997) 658, [arXiv:hep-th/9706109]; M. Aganagic and C. Vafa, [arXiv:hep-th/0012041].}
\REF{\DOUG}{M. Douglas, JHEP {\bf 9707} (1997) 004, [arXiv:hep-th/9612126].}
\REF{\GEO}{C. Vafa, Journ. Math. Phys. {\bf 42} (2001) 2798, [arXiv:hep-th/0008142]; F. Cachazo, K. Intriligator and C. Vafa, Nucl. Phys. {\bf B603} (2001) 3, [arXiv:hep-th/0103067].}
\REF{\SHA}{M. Aganagic, C. Beem and S. Kachru, Nucl. Phys. {\bf B796} (2008) 1, arXiv:0709.4277[hep-th].}
\REF{\NPS}{S. Gukov, C. Vafa and E. Witten, Nucl. Phys. {\bf B584} (2000) 69, Erratum-ibid, {\bf B608} (2001) 477, [arXiv:hep-th/9906070].} 
\REF{\INS}{R. Blumenhagen, M. Cvetic and T. Weigand, Nucl. Phys. {\bf B771} (2007) 113, [arXiv:hep-th/0609191]; L. Ibanez and A. Unranga JHEP {\bf 0703} (2007) 052, 
[arXiv:hep-th/0609213]; B. Florea, S. Kachru, J. McGreevy and N. Saulina, JHEP {\bf 0705} (2007) 024, [arXiv:hep-th/0610003]; L. Ibanez and R. Richter, JHEP {\bf 0903} (2009) 090, 
arXiv:0811.1583[hep-th]; M. Cvetic, J. Halverson and R. Richter, arXiv:0905.3379[hep-th]; D. Green and T. Weigand, arXiv:0906.0595[hep-th]; E. Halyo, arXiv:0906.2159[hep-th]; 
arXiv:0906.2587[hep-th].} 
\REF{\REV}{R. Blumenhagen, M. Cvetic and S. Kachru, T. Weigand, Ann. Rev. Nucl. Part. Sci.(2009) 269, arXiv:0902.3251[hep-th] and references therein.}

\singlespace
\rl{SU-ITP-10-6}
\pagenumber=0
\normalspace
\medskip
\bigskip
\titlestyle{\bf{Metastable Supersymmetry Breaking and Minimal Gauge Mediation on Branes}}
\smallskip
\author{ Edi Halyo{\footnote*{e--mail address: halyo@stanford.edu}}}
\smallskip
\centerline {Department of Physics} 
\centerline{Stanford University} 
\centerline {Stanford, CA 94305}
\smallskip
\vskip 2 cm
\titlestyle{\bf ABSTRACT}

We construct a model with D5 branes wrapped on a deformed and resolved $A_6$ singularity which realizes metastable supersymmetry breaking and minimal gauge mediation.   
Supersymmetry is broken at tree level by the F--term of singlet which also obtains a VEV as required in gauge mediation.
Three nodes of the singularity are used to break supersymmetry whereas the other three realize gauge mediation. The supersymmetry breaking scale is suppressed due to brane instanton 
effects which are computed using a geometric transition.

\singlespace
\vskip 0.5cm
\endpage
\normalspace

\centerline{\bf 1. Introduction}
\medskip

The observable effects of supersymmetry breaking depend on the mechanism that communicates it to the Minimally Supersymmetric Standard Model (MSSM). Gauge mediation[\OGM,\GM,\GUI] is an attractive
mechanism, mainly because when supersymmetry breaking is mediated by MSSM gauge interactions, flavor changing neutral currents are naturally 
suppressed. In this scenario, supersymmetry breaking in the hidden sector is communicated to the observable sector through loop diagrams that contain messenger fields and the MSSM
gauge fields. The main requirement for Minimal Gauge Mediation (MGM) is the existence of a singlet $X$ and messenger fields $(q,\ell),({\bar q}, {\bar \ell})$ in the $5$ and ${\bar 5}$ 
representations of $SU(5)$ (in order to keep gauge coupling unification) with the superpotential
$$W=X(\lambda_{\ell} {\bar \ell} \ell+ \lambda_q {\bar q} q) \eqno(1)$$
In addition, it is assumed that $X$ gets a VEV and a nonzero F--term so that $X=x+\theta^2 F$. Then, the fermionic messengers get masses
$m_q=\lambda_q x$ and $m_{\ell}=\lambda_{\ell} x$.
Due to supersymmetry breaking, the scalar messengers are split from the fermionic ones and have masses squared (where $F<x^2$ is assumed to avoid tachyons)
$$m_q^2=\lambda_q^2 x^2 \pm \lambda_q F \qquad m_{\ell}^2=\lambda_{\ell}^2 x^2 \pm \lambda_{\ell} F \eqno(2)$$

This mass splitting is communicated to the superpartners through the gauge interactions giving rise to soft supersymetry breaking masses. Gaugino masses arise at one--loop with 
$$m_{\lambda_i}={\alpha_i \over \pi} \Lambda \eqno(3)$$
whereas the soft masses squared for squarks and sleptons arise at two loops with
$$m_s^2=2 \Lambda^2 \left[C_3 \left({\alpha_3 \over {4 \pi}}\right)^2+ C_2 \left({\alpha_2 \over {4 \pi}}\right)^2+ {5 \over 3}({Y \over 2})\left({\alpha_1 \over {4 \pi}}\right)^2 
\right] \eqno(4)$$
where $\Lambda=F/x$ and $C_{1,2,3}$ are group theoretic factors for the three MSSM gauge groups. $\Lambda$ is the only physical parameter since it
determines all the soft masses; therefore $x$ and $F$ can be both quite large or small. (However,
$x>TeV$ since the messengers have not been observed.) Due to eqs. (3) and (4), TeV scale soft masses require $\Lambda \sim 100~TeV$.

Aa mentioned above, gauge mediation of supersymmetry requires a singlet with a VEV and a nonzero F--term. This has been difficult to obtain in most models of dynamical supersymmetry 
breaking[\SUSY] mainly because these are models of strongly interacting chiral gauge theories (with no singlets). Recently,
it was realized that supersymmetry may be broken in nonsupersymmetric metastable vacua, i.e. in supersymmetric models with local nonsupersymmetric minima of the scalar potential[\ISS]. 
This new possibility greatly facilitates the construction of models with a singlet like $X$ above. At these false vacua, supersymmetry can be broken at tree level by the nonzero F--term of 
a singlet. The necessary suppression of the supersymmetry scale can be obtained by retrofitting the dimensionful parameters of the models[\DIN] rather than by dynamical effects.
Unfortunately, in most models of metastable supersymmetry breaking, the field that breaks supersymmetry does not obtain a VEV, i.e. the metastable vacuum is at the origin
[\ISS,\RAY-\ESUS]{\footnote 1{However see ref. [\RSY] for an example with nonzero VEVs.}}. 
In this case, gauge mediation cannot be realized since the singlet VEV is necessary to give large masses to the messengers. Thus, finding a model of gauge mediation is not trivial
even in the context of metastable supersymmetry beaking.

In addition to field theories, metastable supersymmetry breaking and mediation have been studied in brane models[\BGH-\LAST]. Most known and some new models of supersymmetry breaking 
have been obtained on branes.
In this context, it was noticed that quiver models obtained from branes wrapped on singularities naturally have singlets with Yukawa couplings like the one
in eq. (1)[\KAW]. In particular, models with D5 branes wrapped on $A_n$ type singularities may naturally realize gauge mediation. This is due to the fact that in these brane constructions, 
each node of the singularity gives rise to a
(singlet if there is only one D5 brane wrapped on the node) scalar field. This singlet couples to its two neighboring nodes through Yukawa--like couplings to pairs of fields in the 
bifundamental representations of the gauge groups. Thus, this singlet can serve as the $X$ field in eq. (1). On one side it is coupled to the hidden supersymmetry 
breaking sector and on the other side to the messenger fields by Yukawa--like couplings. 

As a result, we can build models on an $A_n$ singularities in which a number of nodes realize supersymmetry breaking, the next node
contains the $X$ field above, which couples to the bifundamental messenger fields. These in turn couple to the MSSM fields through their common gauge interactions.
The quiver has a modular structure in which $X$ lives on a particular node. The supersymmetry breaking hidden sector lives on the nodes to (say) the left of this node whereas 
the messenger fields and MSSM live on the nodes to its right..

In this paper, we construct a brane model that breaks supersymmetry in a metastable vacuum and communicates this to the MSSM by gauge mediation. The model is obtained on the world--volumes
of D5 branes that are wrapped on an $A_6$ singularity fibered over a complex plane $C(x)$. Different nodes of the singularity are deformed and resolved; in addition 
all nodes except for two have large volumes (compared to the string scale). The relatively large number of nodes of the singularity is due to complexity of the model that both realizes and 
mediates supersymmetry breaking.
Three nodes are needed for supersymmetry breaking, one node contains the $X$ field and the messengers and MSSM live on the remaining two nodes. 
We describe in detail the function of each node in the construction and show how the resulting world--volume theory leads to metastable supersymmetry breaking and gauge mediation.
At low energies, the world--volume field theory we obtain is a slight variation of that in ref. [\DIN].
In the metastable vacuum supersymmetry is broken by both F and D--terms at tree level; however only the F--term breaking is relevant for gauge mediation. The supersymmetry breaking scale 
is suppressed due to a D1 brane instanton effect which plays the role of retrofitting in field theory.

This paper is organized as follows. In the next section, we review the basics of the brane construction on $A_n$ singularities and describe the $A_6$ model in detail. In setion 3, we
analyze the world-volume field theory and show that it breaks supersymmetry in a metastable vacuum and mediates this to MSSM through gauge interactions. Section 4 contains our conclusions
and a discussion of our results.

\bigskip
\centerline{\bf 2. The Brane Model on an $A_6$ Singularity}
\medskip

In this section, we construct a brane model that realizes metastable supersymmetry breaking and gauge mediation. First, we briefly review the relations between the
properties of $A_n$ singularities and the world--volume theories of D5 branes wrapped on them. We then describe the model with D5 branes wrapped on an $A_6$ singularity, including
the contributions of each node separately. 

{\bf 2.1. World--volume theories of D5 branes wrapped on $A_n$ singularities:} Consider an $A_n$ singularity fibered over the complex plane $C(x)$ given by
$$uv=(z-z_1(x))(z-z_2(x)) \ldots (z-z_n(x))(z-z_{n+1}(x)) \eqno(5)$$
where $z_i(x)$ parametrize the fibering of $A_n$ over $C(x)$. This singularity has $n$ nodes ($S^2$s). If we wrap $N_i$ D5 branes on the $i$th node, we get the product gauge group 
$\Pi_i U(N_i)$ in the world--volume theory. The gauge couplings are given by[\QUI]
$${{4 \pi} \over g_i^2}={V_i \over {(2\pi)^2 g_s \ell_s^2}} \qquad i=1,2, \ldots,n \eqno(6)$$
where $g_s$ and $\ell_s$ are the string coupling and length respectively and $V_i$ is the stringy volume of the $i$th node given by $V_i=(2 \pi)^4 \ell_s^4(B_i^2+r_i^2+\alpha_i^2)^{1/2}$.
Here
$$B_i=\int_{S_i^2}B^{NS} \qquad r_i^2=\int_{S_i^2}J  \eqno(7)$$
i.e. $B_i$ is the NS--NS flux through the $i$th node and $r_i^2$ is the volume of the blown--up $S_i^2$s. The world--volume matter sector[\QUI] contains fields in the adjoint or fundamental
representations of the gauge groups. On the $i$th node, there is a field $\phi_i$ that is an adjoint of $U(N_i)$. Between any
two neighboring nodes, $i$ and $j=i+1$ there is a pair of fields $Q_{ij},Q_{ji}$ in the bifundamental representations of $U(N_i) \times U(N_j)$. The superpotential contains Yukawa couplings
between these fields[\QUI,\SUP]
$$W=\sum_{ij} Q_{ij}Q_{ji}(\phi_{i+1}-\phi_i) \eqno(8)$$
In addition, the deformation and fibering of the singularity lead to superpotential terms of the type
$$W(\phi_i)=\int^{\phi_i}(z_{i+1}(x)-z_i(x)) dx  \eqno(9)$$
Resolving (blowing--up) the $i$th node so that $r_i^2 \not =0$, introduces an anomalous D--term, $\xi_i$[\DOUG], for the gauge group $U(1)_i$ (or the Abelian subgroup of $U(N_i)$).
Also, a geometric transition may take place on any of the nodes leading to $S^2 \to S^3$ on the node. After the transition, the D5 branes wrapped on this node are replaced by 
Ramond--Ramond flux.
Geometric transitions are particularly useful for computing nonperturbative effects on a node such as those of a D1 brane instanton. The model we describe below contains both resolutions 
and geometric transitions. 

{\bf 2.2. The model with D5 branes on an $A_6$ singularity:} We now consider D5 branes wrapped on the $A_6$ singularity defined by
$$\eqalignno{
uv=z(z-m(x-a))&(z+m(x+a))(z-m(x-a))(z-m(x+a)+\lambda x^2) \cr
                                  &(z+m(x-a)+\lambda x^2)(z-ma+\lambda x^2) &(10) \cr}$$
This singularity has six nodes; we wrap $5$ D5 branes on the fifth node and one D5 brane on each of the others. This leads to the gauge group $U(1)_1 \times U(1)_2 \times U(1)_3 \times
U(1)_4 \times U(5) \times U(1)_6$ on the world--volume. We decouple all the gauge groups except $U(1)_3$ (which will acquire an anomalous D--term) and $U(5)$ which contains the MSSM
gauge groups. The decoupling of these gauge interactions is accomplished by taking the corresponding stringy volumes of the nodes to be very large, e.g. by assuming large NS--NS fluxes 
$B_i>>1$ on the nodes. Then, by eq. (6)
we find that the gauge couplings are very small and these symmetries become global. In addition, we decouple all heavy fields with mass $\sim m \sim a \sim M_s$ and ignore the norenormalizable
terms they induce in the superpotential since these do not affect the physics at low energies relevant for supersymmetry breaking.

We can find the superpotential that arises from the singularity using the information from the previous subsection. Below we list the contribution of each node to matter and the superpotential
separately and discuss their roles in metastable supersymmetry breaking and gauge mediation. The first node contains the fields $\phi_1, Q_{12},Q_{21}$ with the superpotantial
$$node~1: \qquad W_1={m \over 2}(\phi_1-a)^2-\phi_1 Q_{12}Q_{21} \eqno(11)$$
This node is used for generating an exponentially suppressed scale through D1 brane instanton effects. The brane instanton contribution to the superpotential is an F--term and can be computed
after the node goes through a geometric transition in which $S^2 \to S^3$. This effect is calculable only if the node is isolated and all the fields living on it are massive.
We see from eq. (11) that the first node is isolated at $x=a$ and $\phi_1,Q_{12},Q_{21}$ are massive. Therefore, we can use a geometric transition[\GEO] to obtain the 
brane instanton induced F--term.
After the geometric transition, the D5 brane wrapped on this node is replaced by Ramond--Ramond flux $\int_{S^3} H^{RR}=1$ and $\phi_1,Q_{12},Q_{21}$ disappear from the spectrum. The 
geometry becomes[\SHA]
$$\eqalignno{
uv=(z-s)(z-m(x-a))&(z+m(x+a))(z-m(x-a))(z-m(x+a)+\lambda x^2) \cr
                                  &(z+m(x-a)+\lambda x^2)(z-ma+\lambda x^2) &(12) \cr}$$
where $s=mS$ and $S$ is the size of the resolved (blown--up) $S^3$. Due to the transition, there are two new contributions to $W$. The first is the flux superpotential[\NPS]
$$W_{flux}={V_1 \over {2 \pi g_s \ell_s}}S+S \left(log{S \over \Delta^3}-1 \right) \eqno(13)$$
The second is the instanton correction to the $\phi_2$ superpotential given by[\SHA]{\footnote2{For some examples of brane instanton effects see refs.[\INS,\REV].}}
$$W^{\prime}(\phi_2)=\int^{\phi_2}(z_2-{\bar z_1}) dx \eqno(14)$$
where ${\bar z_1}$ is the solution to $z(z-m(x-a))=s $
that is asymptotic to $z_1$. We find
$$W^{\prime}(\phi_2)=-{S \over 2} log{{|\phi_2-a|} \over \Delta} \eqno(15)$$
which gives rise to the F--term for $\phi_2$ 
$$F_{\phi_2}={\partial W^{\prime} \over \partial \phi_2}=-{S \over {2a}} \eqno(16)$$
Decoupling $S$ by setting its F--term to zero fixes its VEV which is exponentially suppressed
$$S=S_0=\Delta^3 e^{-V_1/2 \pi g_s \ell_s^2} \eqno(17)$$
Thus brane instanton effects on the first node lead to an exponentially suppressed F--term for $\phi_2$
$$F_{\phi_2}=-{S_0 \over {2a}}=-{\Delta^3 \over {2a}}e^{-V_1/2 \pi g_s \ell_s^2} \eqno(18)$$

This in turn results in exponentially small masses for the bifundamentals $Q_{23},Q_{32}$ which live on the second node. These fields with small masses appear in the $U(1)_3$ 
D--terms which includes an anomalous D--term. They effectively cause the nonzero D--term to
be exponentially small rather than about $\xi \sim M_s^2$. This is the origin of the exponentially small supersymmetry breaking scale in our model.  

On the second node we find the fields $\phi_2, Q_{23},Q_{32}$ with 
$$node~2: \qquad W_2=-m\phi_2^2+ \phi_2(Q_{12}Q_{21}-Q_{23}Q_{32}) \eqno(19)$$
In addition, we have the exponentially small F--term, $F_{\phi_2} \phi_2$ where $F$ is given by eq. (18). The field $\phi_2$ decouples at low energies since it has a large mass $m \sim M_s$.
Setting its F--term to zero we find that $\phi_2$ obtains an exponentially small VEV, $\phi_2=F_{\phi_2}/2m$ which gives small masses to $Q_{23},Q_{32}$
$$m_Q= {\Delta^3 \over {4ma}}e^{-V_1/2 \pi g_s \ell_s^2} \eqno(20)$$
This the overall contribution of the second node. 

The third node on which $\phi_3,Q_{34},Q_{43}$ live contributes
$$node~3: \qquad W_3=m\phi_3^2+\phi_3(Q_{23}Q_{32}-Q_{34}Q_{43}) \eqno(21)$$
In addition, we blow--up the third node by taking $r_3^2 \not =0$ which gives rise to an anomalous $U(1)_3$ D--term, $\xi$, so that
$$D_3=|Q_{23}|^2-|Q_{32}|^2-|Q_{34}|^2+|Q_{43}|^2- \xi  \eqno(22)$$
As mentioned above, the small masses for $Q_{23},Q_{32}$ are the origin of the small nonzero D--term after minimizing the scalar potential and give rise to exponentially suppressed supersymmetry
breaking. $\phi_3$ is heavy and decouples with vanishing VEV.

On the fourth node we find the fields $\phi_4,Q_{45}Q_{54}$ with
$$node~4: \qquad W_4=2ma \phi_4-{\lambda \over 3} \phi_4^3+\phi_4(Q_{34}Q_{43}-Q_{45}Q_{54}) \eqno(23)$$
Contrary to the other singlets, the field $\phi_4$ does not have a mass term and does not decouple. It has an F--term and a nontrivial potential. We will see that it plays the role of
$X$ in eq. (1). The F--term and VEV of $\phi_4$ are the main ingredients of gauge mediation. Supersymmetry breaking is communicated to $\phi_4$ by the fields $Q_{34},Q_{43}$. The fields
$Q_{45},Q_{54}$ are the messengers which get masses from the VEV of $\phi_4$. All the fields in this node remain in the low energy spectrum and play crucial roles in supersymmetry breaking and
gauge mediation.

There are $5$ D5 branes wrapped on the fifth node giving rise to a $U(5)$ gauge group that contains the MSSM gauge group. In addition, there are the bifundamentals $Q_{56},Q_{65}$ which
are the MSSM matter fields (which are massless at this level) in the $5$ and ${\bar 5}$ representations just like the messengers $Q_{45},Q_{54}$. 
The superpotential arising from this node is
$$node~5: \qquad W_5=-m \phi_5^2+\phi_5(Q_{45}Q_{54}-Q_{56}Q_{65}) \eqno(24)$$
Again $\phi_5$ decouples since it is heavy. This description of 
MSSM is too simple and not very realistic; however, it can be improved by combining the fourth node to a more complicated singularity on which MSSM lives. The modular nature of our 
construction easily allows this possibility.

Finally, from the sixth node we get
$$node~6: \qquad W_6={m^2 \over 2} \phi_6^2+\phi_6 Q_{56}Q_{65} \eqno(25)$$
$\phi_6$ is massive and decouples with vanishing VEV. It gives rise to dimension five and six nonrenormalizable terms containing MSSM fields. These terms may lead to baryon and lepton number 
violation or flavor changing neutral currents. However, since the MSSM sector of our model is not very realistic we will not worry about this problem and assume that a proper construction
of the MSSM sector will suppress these operators.

\bigskip
\centerline{\bf 3. Supersymmetry Breaking and Gauge Mediation}
\medskip

We now show that the above brane configuration leads to supersymmetry breaking in a metastable vacuum and its gauge mediation to the observable sector. This analysis closely follows that in 
ref. [\DIN] with minor differences. At low energies, i.e $E<<M_s$, the total superpotential obtained from the above brane construction is (leaving out MSSM)
$$W=\phi_4(Q_{34}Q_{43}-Q_{45}Q_{54})+2ma\phi_4-{\lambda \over 3} \phi_4^3+m_Q Q_{23}Q_{32} \eqno(26)$$
The F--terms obtained from $W$ are
$$F_{\phi_4}=Q_{23}Q_{32}-Q_{34}Q_{43}+2ma-\lambda \phi_4^2 \eqno(27)$$
$$F_{Q_{23}}=m_Q Q_{32} \qquad F_{Q_{32}}=m_Q Q_{23} \eqno(28)$$
$$F_{Q_{34}}=\phi_4 Q_{43} \qquad F_{Q_{43}}=\phi_4 Q_{34} \eqno(29)$$
$$F_{Q_{45}}=-\phi_4 Q_{54} \qquad F_{Q_{54}}=-\phi_4 Q_{45} \eqno(30)$$ 
The scalar potential is given by $V_F=\sum |F_i|^2$.
In addition, there is the $U(1)_3$ D--term contribution to the scalar potential
$$V_D=g_3^2(|Q_{23}|^2-|Q_{32}|^2-|Q_{34}|^2+|Q_{43}|^2- \xi)^2  \eqno(31)$$
The total scalar potential is $V=V_F+V_D$.

We now specialize to the case with $Q_{34}=Q_{43}=Q_{45}=Q_{54}=0$. Minimizing the scalar potential with respect to $Q_{23}$ and $Q_{32}$ we find that (for $g_3^2 \xi>m_Q^2$ which holds in 
our case)
$$|Q_{23}|^2=\xi-{m_Q^2 \over g_3^2} \qquad Q_{32}=0 \eqno(32)$$
Then, the nonzero D--term effectivelybecomes $|D|=m_Q^2/g_3^2$ which is exponentially suppressed due to eq. (31). This is the main contribution of the fields $Q_{23},Q_{32}$ to 
supersymmetry breaking. Even though they have small masses and remain in the low--energy spectrum, they decouple from the supersymmetry breaking physics beyond the above effect. Therefore we 
can neglect them below. The low energy scalar potential becomes (excluding the $Q_{23},Q_{32}$ dependent terms and a constant term)
$$V=|2ma- \lambda \phi_4^2+Q_{34}Q_{43}|^2+|\phi_4|^2(|Q_{34}|^2+|Q_{43}|^2)+g_3^2(-|Q_{34}|^2+|Q_{43}|^2+{m_Q^2 \over g_3^2})^2  \eqno(33)$$
We see that the scalar masses squared for $Q_{34},Q_{43}$ are $m_s^2=|\phi_4|^2 \pm m_Q^2/g_3^2$ whereas their fermionic partners have masses $m_f=|\phi_4|$. 
There are mixing terms for the scalars coming from the first term in eq. (33) and a nonzero F--term for $\phi_4$
(which we obtain below). We neglect these contributions since as we show below they are small, $O(m_Q^4/ma)<<O(m_Q^2)$.
Due to the above splitting of masses, $\phi_4$ obtains a one--loop potential given by
$$V_1={m_Q^4 \over {16 \pi^2 g_3^2}} log \left({|\phi_4|^2 \over \Lambda^2}\right) \eqno(34)$$
The total potential for $\phi_4$ has a metastable minimum at
$$|\phi_4|^2={{ma} \over {\lambda}} \left(1+\sqrt{1-{m_Q^4 \over {32 \pi^2 m^2a^2}}} \right) \simeq {{2ma} \over \lambda} -{m_Q^4 \over {64 \pi^2 m a}} \eqno(35)$$
In this vacuum we find that supersymmetry is broken due to the nonzero F--term
$$F_{\phi_4}={m_Q^4 \over {128 \pi^2 m a}} \eqno(36)$$

The nonzero VEV and F--term for $\phi_4$ lead to the mass and mixing terms for the messenger fields $Q_{45},Q_{54}$ and induce a mass splitting just like in the MGM mechanism in eqs. (2). 
These in turn give rise to soft supersymmetry breaking masses for the MSSM fields, $Q_{56},Q_{65}$, as in eqs. (3) and (4). For successful gauge mediation without tachyons we need 
$F_{\phi_4}<\phi_4$ which is the case since the F--term is proportional to $m_Q^2$ which is exponentially suppressed. We find that $\Lambda=F_{\phi_4}/\phi_4$ is
$$\Lambda \simeq {\lambda \over {256 \pi^2}} {m_Q^4 \over {(ma)^{3/2}}} \eqno(37)$$
which can be $O(100~TeV)$ for a choice of $m \sim a \sim M_s \sim 10^{17}~GeV$ and $m_Q \sim 10^{15}~GeV$. The two orders of magnitude suppression of $m_Q$ is easily obtained from eq. (20).

Note that supersymmetry is broken not only by $F_{\phi_4} \not =0$ but also by the nonzero F and D--terms 
$$F_{Q_{32}}=m_Q Q_{23} \sim m_Q \sqrt{\xi} \qquad D={m_Q^2 \over g_3^2}  \eqno(38)$$
In the limit of global supersymmetry (with gravity decoupled) which is obtained with a noncompact space transeverse to the branes, these cannot be communicated to the
observable sector. In principle, they may contribute to soft masses due to nonrenormalizable terms which connect the second and third nodes (with the above nonzero F and D--terms)
to the fifth node on which MSSM fields live. However, in the above model, these are either too small or zero due to vanishing VEVs. Therefore, even though, in this model supersymmetry
breaking occurs through two F and one D--term that are nonzero only $F_{\phi_4}$ is phenomenologically relevant since it is the only effect coupled to the MSSM.

\bigskip
\centerline{\bf 4. Conclusions and Discussion}
\medskip

In this paper, we constructed a model with D5 branes wrapped on a deformed, resolved and fibered $A_6$ singularity. The model
breaks supersymmetry in a metastable vacuum at tree level by nonzero F and D--terms. The supersymmetry breaking scale is suppressed due to
brane instanton effects which are calculated through a geometric transition. Supersymmetry breaking is then mediated to the MSSM sector through gauge interactions. The quiver theory
that results has a modular structure. Three nodes realize supersymmetry breaking, one node contains the singlet (that couples to the messengers) and the messengers and MSSM live
on the remaining two nodes.

One of the main shortcomings of the model is the realization of the MSSM sector. The MSSM fields live on a node with a $U(5)$ gauge group and matter is in the 
$5$ and ${\bar 5}$ representations which is not realistic because the $10$ representation is missing. This problem is related to the chiral nature of MSSM 
matter. In general, chiral matter can be obtained on $A_n$ singularities only if they are orbifolded or orientifolded. Therefore, a more realistic MSSM sector requires 
a more complicated singularity. 

Another potential problem arises if the model is realized on compact transverse space so that gravity is not decoupled. In this case, the nonzero F and D--terms in eq. (38) cannot be
neglected since these effects can be communicated to the visible sector by gravity. For the above choice of parameters which leads to phenomenologically acceptable gauge mediation, we find
that gravity mediated contributions to soft masses are too large, e.g. $\sim 10^{15}~GeV$. One possible solution is to make exponentially suppressed mass $m_Q$ very small e.g. $O(TeV)$.
Then the gravity mediated contributions to soft masses are also of $O(TeV)$ but the gauge mediated contributions are completely negligible. Of course, this introduces the flavor changing 
neutral current problem
which is the main motivation to consider gauge mediation in the first place. Thus, we conclude that the above model has problems that are hard to avoid if gravity is not decoupled. However,
this does not mean that all models obtained by wrapping D5 branes on $A_n$ singularities share the same problem. It would be worthwhile to try to build a model with only one nozero F--term
as a source of supersymmetry breaking which avoids the problems arising from gravity mediation.

In field theory the problem with gravity mediation can be solved by taking the parmeters $m,a$ and $\xi$ to be relatively small (compared to the string scale) by retrofitting. 
Then, acceptable gauge mediation can be obtained without large contributions from gravity mediation. In string theory, such small values are unnatural. On the other hand, small values 
of $m,a$ can be obtained by instanton effects just like for $m_Q$ above. This requires two more D1 brane instanton effects calculated through two more geometric transitions on two new 
nodes. Clearly, the model then becomes quite complicated and unattractive.

Minimal gauge mediation can be generalized in different ways. First, there can be more than one field like $X$ in eq. (1) which leads to soft masses for squarks and sleptons that are more
general than those in eqs. (3) and (4).
Second, the messenger sector can be generalized to include more than one pair of $5,{\bar 5}$. Unfortunately, quiver models obtained on world--volumes of D5 branes wrapped on
$A_n$ singularities are not easily amenable to the above generalizations. The only possibility seems to be placing $N$ D5 branes on the node that gives the field $X$ (or $\phi_4$). This
gives rise to an adjoint instead of the singlet. In addition, now, the messengers are also in the fundamental representation of this new $U(N)$. Even if we decouple the $U(N)$ by taking its
coupling to zero, the number of ``singlets'' and messenger pairs is always the same with the (now) global $U(N)$ symmetry guaranteeing the equality of Yukawa couplings between the
``singlet'' and the messengers. This does not quite lead to general gauge mediation.
It would be interesting to obtain general gauge mediation models on branes wrapped on singular spaces more complicated than $A_n$.

\bigskip
\centerline{\bf Acknowledgements}

I would like to thank the Stanford Institute for Theoretical Physics for hospitality.

\vfill

\refout

\end
\bye